\documentclass[10pt]{article}

\usepackage{amsmath}
\usepackage{amssymb}

\usepackage{graphicx}
\usepackage{placeins}
\usepackage{longtable}

\usepackage{cite}

\usepackage{color}

\usepackage[labelfont=bf,labelsep=period,justification=raggedright]{caption}

\date{}

\begin{document}

\begin{flushleft}
{\Large
\textbf{Information Evolution in Social Networks}
}

Lada A. Adamic$^{1,2,\ast}$, 
Thomas M. Lento$^{2}$, 
Eytan Adar$^{1}$,
Pauline C. Ng$^{3}$
\\
\bf{1} School of Information, University of Michigan, Ann Arbor, MI, CA
\\
\bf{2} Facebook, Menlo Park, CA, USA
\\
\bf{3} Genome Institute of Singapore
\end{flushleft}

\section*{Abstract}

Social networks readily transmit information, albeit with less than perfect fidelity.  We present a large-scale measurement of this imperfect information copying mechanism by examining the dissemination and evolution of thousands of memes, collectively replicated hundreds of millions of times in the online social network Facebook. The information undergoes an evolutionary process that exhibits several regularities. A meme's mutation rate characterizes the population distribution of its variants, in accordance with the Yule process. Variants further apart in the diffusion cascade have greater edit distance, as would be expected in an iterative, imperfect replication process. Some text sequences can confer a replicative advantage; these sequences are abundant and transfer ``laterally'' between different memes. Subpopulations of the social network can preferentially transmit a specific variant of a meme if the variant matches their beliefs or culture. Understanding the mechanism driving change in diffusing information has important implications for how we interpret and harness the information that reaches us through our social networks.

\section*{Introduction}

Richard Dawkins coined the word ``meme'' to designate ideas or messages that spread and evolve analogously to genes through communication\cite{dawkins1976selfish}. Determining the extent to which the gene analogy applies to memes has been hampered by the lack of large-scale data containing the evolution histories of many memes, where one can pinpoint when and where information was modified and how the modifications were subsequently diffused. In this study we are able to gather precisely such data, consisting of near-complete traces of thousands of memes, collectively comprising over 460 million individual instances propagated via Facebook. Facebook is an online social network that enables individuals to communicate with their friends.  The primary mechanism for this communication are {\it status updates} which appear in news feeds of friends. A message or idea that is appealing can quickly propagate, as those friends can choose to post the message as their own status update, thus exposing their own friends. In this way Facebook acts as a large petri dish in which memes can mutate and replicate over the substrate of the network of friendship ties. While there are other environments where memes flourish, those memes that do enter Facebook can be examined in detail, uncovering mechanisms previously difficult--or impossible--to study.

We analyzed anonymized data gathered over a period of 18 months, from April 2009--October 2011 during which many memes were propagated as textual status updates on Facebook.   In April 2009 the character limit on the status update had been increased, allowing one to not only express a complete idea or story, but also add replication instructions, e.g. ``copy and paste'' or ``repost''. Widely propagated memes would typically carry these instructions because there was yet no ``share'' functionality, which could at the click of a button exactly replicate the information as one's own update. As a result, memes propagating via a manual copy and paste mechanism can be exact, or they might contain a ``mutation", an accidental or intentional modification. Since the notion of a meme was first based on the parallels between genes and ideas, we consider how textual status updates match genes in their structure and mechanics. First, they encode information, whether it is a joke, a warning, or a call to action, in a way that parallels genetic information: a string of characters is ``transcribed", allowing for some characters to be added, deleted, or substituted. Second, the replication (accurate or inaccurate) can be performed by anyone exposed to the meme. There are other forms in which memes occur in online environments, e.g. photographs or videos. However, the information in these is not as readily analyzed, and they cannot be as easily modified by anyone as spoken or written ideas can. We therefore focus our attention on textual status updates on Facebook to understand how information evolves when anyone can easily modify and retell it.   

\begin{figure}[tbh]
\centering
\includegraphics[width=0.8\columnwidth]{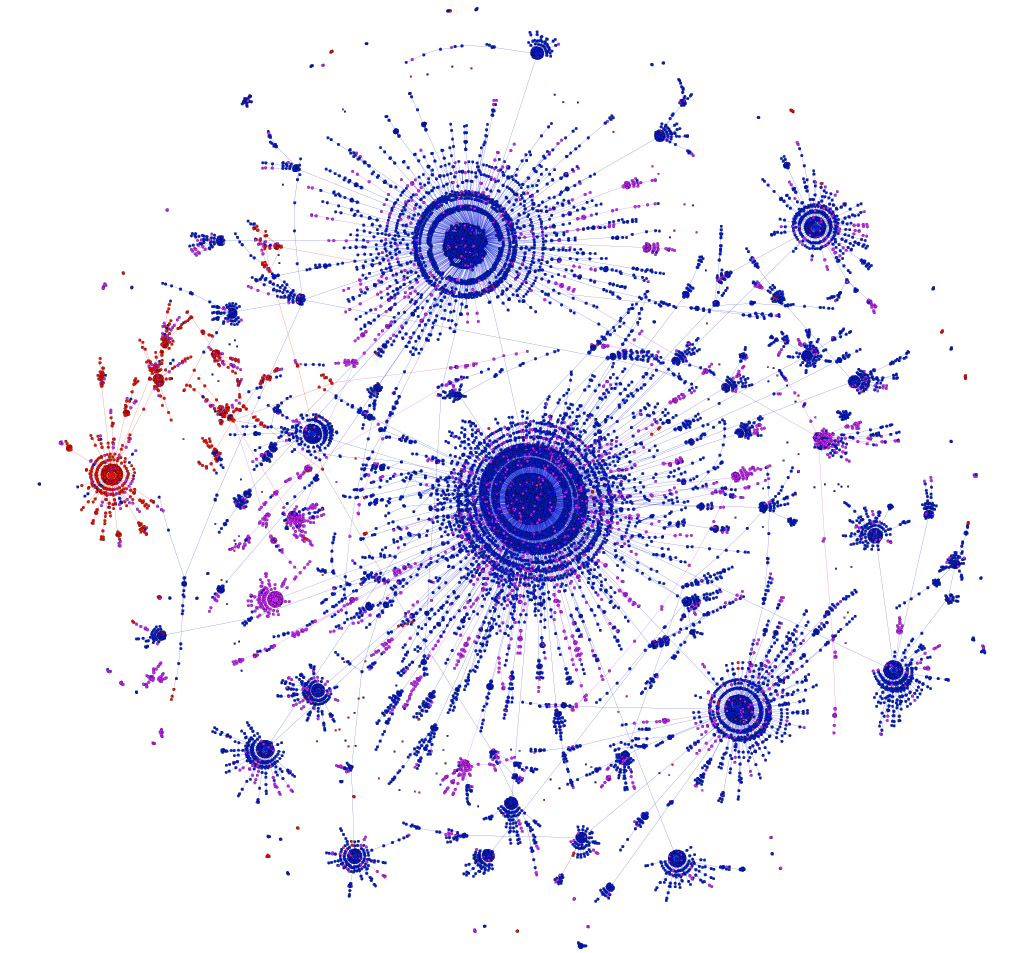}
\caption{{\bf Approximate phylogenetic forest of the ``no one should'' meme.} Each node is a variant, and each edge connects a variant to the most likely ancestor variant. Nodes are colored by timing prompt: rest of the day (blue), next 24 hours (red), or other (purple), showing that mutations in the timing prompt are preserved along the branches of the tree. \label{fig:nooneshouldphylo}}
\end{figure}

Prior studies have either analyzed the overall popularity of memes~\cite{leskovec2009meme,Ratkiewicz11meme,simmons2011memes,romero2011differences,weng2012competition,coscia2013memes}, or examined a small sample of meme variants in detail~\cite{heath2001emotional,bennett2003chain,liben2008tracing,shifman09meme,shifman2011anatomy}. However, neither the large-- nor small--scale studies have been able to formulate a model by which new meme variants arise. This has left a gap in our understanding of the mechanism by which social networks can cause information to evolve, a mechanism that carries important implications for the fidelity of all socially transmitted information, as well as for specific types of information such as political campaign messages~\cite{Ratkiewicz11meme}. It is also unclear the extent to which the biological analogy of genetic evolution carries over into the evolution of memes. We will use the terminology from genetics in describing the processes memes undergo, but defer evaluation of the biological analogy to the Discussion.

\section*{Results}

\subsection*{Replication}   
In this study, we are interested in memes that are replicated when a user copies a piece of text from a friend. Typically, this is done by
selecting the text of the friend's update, copying it, pasting it into one's own status update box, and sometimes editing the text further before clicking `post'. However, the abundance
of typos in some variants indicates that they were retyped character-by-character as opposed to replicated via a block copy-and-paste action. 
We define the mutation rate $\mu$ as the proportion of copies which introduce new edits as opposed to creating exact replicas. We further link each
new variant to the variant it was most likely derived from, as described below.

To see concretely how
mutation occurs we consider the meme ``No one should die because
they cannot afford health care and no one should go broke because they
get sick. If you agree please post this as your status for the rest
of the day.'' In this form, which emerged early on in the meme's
evolution, it was copied over 470,000 times. A variant prepending
``thinks that'' (which would follow the individual's name), was copied
60,000 times. The third most popular variant inserted ``We are only as
strong as the weakest among us'' in the middle. Some copies changed 
``the rest of the day'' to ``the next 24 hours.'' Figure~\ref{fig:nooneshouldphylo} shows how this particular timing trait is passed down through the genetic lineage.  The high number of 
exact copies is consistent with this meme having a modest mutation rate
$\mu =  0.11$. That is, 89\% of the copies were exact, while 11\% introduced
a mutation.

\begin{figure}[tbh]
\centering
\includegraphics[width=0.75\columnwidth]{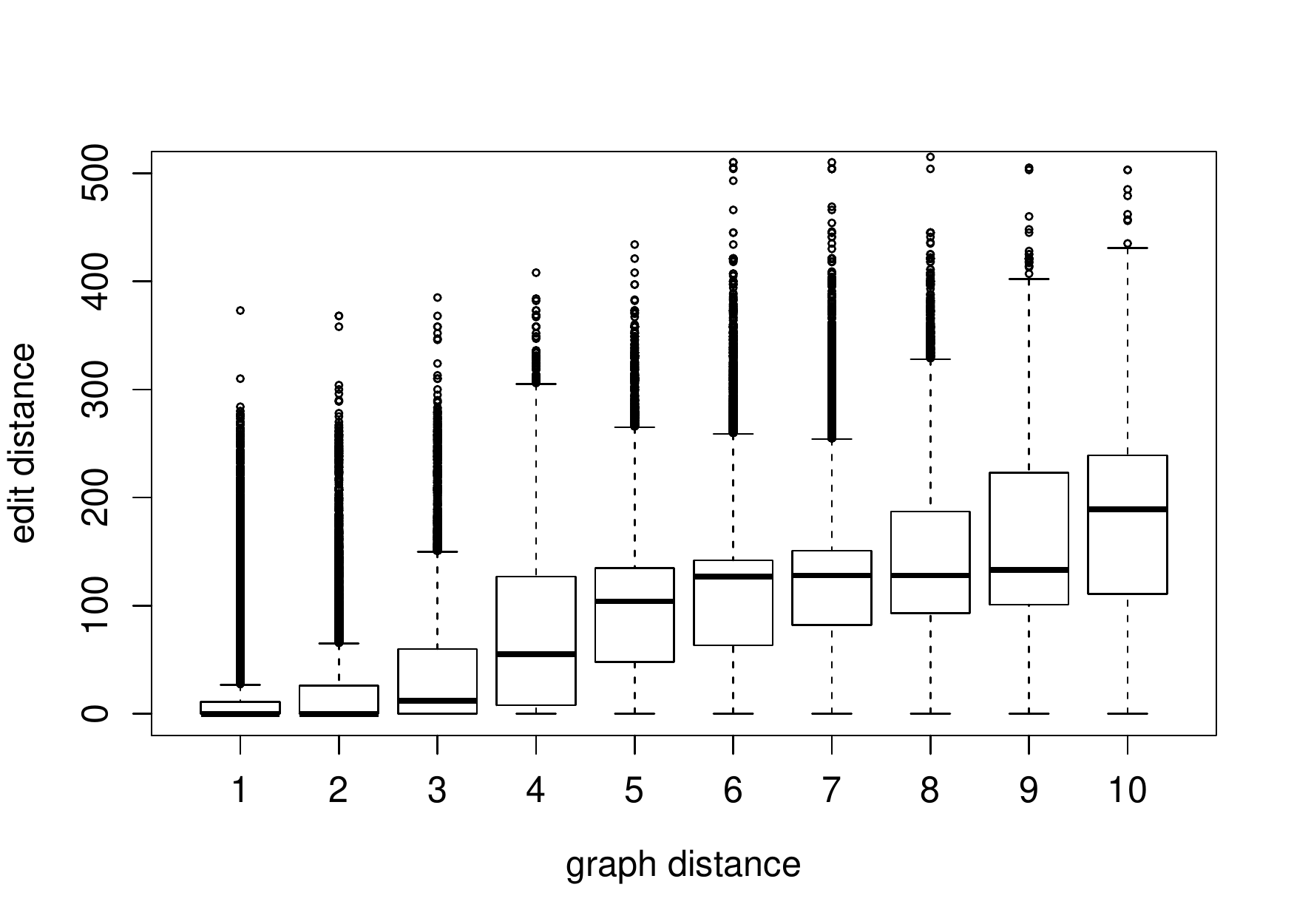}
\caption{{\bf Edit distance from the root typically increases as a node is further down in the copy chain, i.e. the greater the number of hops in the diffusion tree).} Here edit distance is shown for a sample of memes where copy chains reached a length of 10 or greater.\label{fig:editgraphdistance}} 
\end{figure}

In order to trace the lineage of particular variants, we use the edit 
distance of between the variant a user posted and the variants previously posted by the user's friends. 
We define the edit distance as the number of character
additions and deletions that must be performed in order to
obtain one variant of the meme from another. For example, changing
the phrase ``thinks that'' to ``agrees that'' creates two variants
separated by edit distance 10,
since the 5 characters `t h i n k' are deleted and `a g r e e' are added. 
As might be expected, there is a compound effect of such mutations over several generations,
with edit distance increasing as copy is made from copy is made 
from copy
(Figure~\ref{fig:editgraphdistance}).

Before examining the lineage of any particular copy,
we cluster variants into memes according to their textual similarity (see {\it Materials \& Methods} for details) in order to identify the memes captured in the data.
This processes yields a set of 4087 large memes, each of which has at least one variant with 100 copies between April 2009 and October 2011.

\subsection*{Population Genetics of Memes} 
\begin{figure}[tbh]
\centering
\begin{tabular}{cc}
\includegraphics[width=0.5\columnwidth]{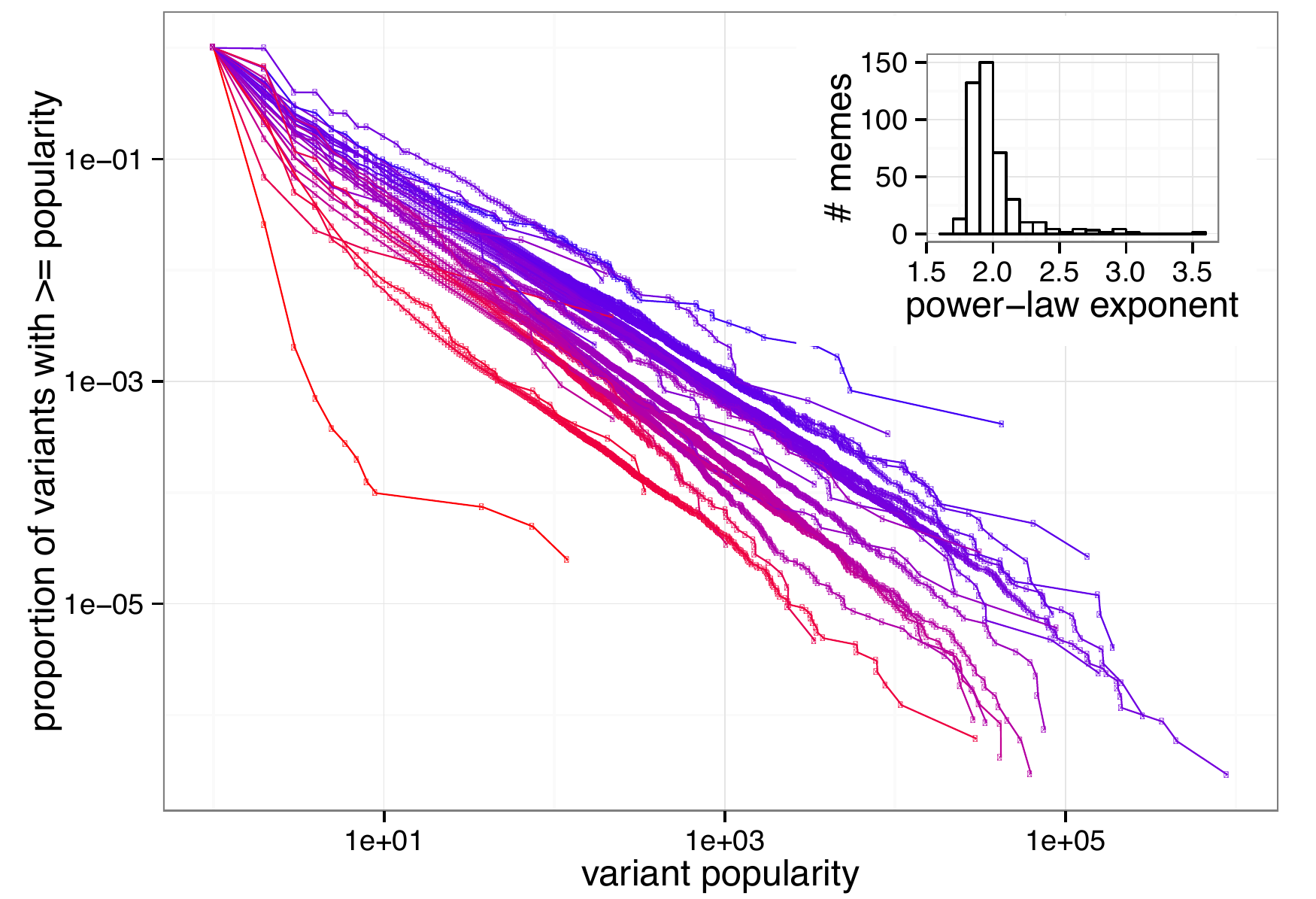}  &
\includegraphics[width=0.5\columnwidth]{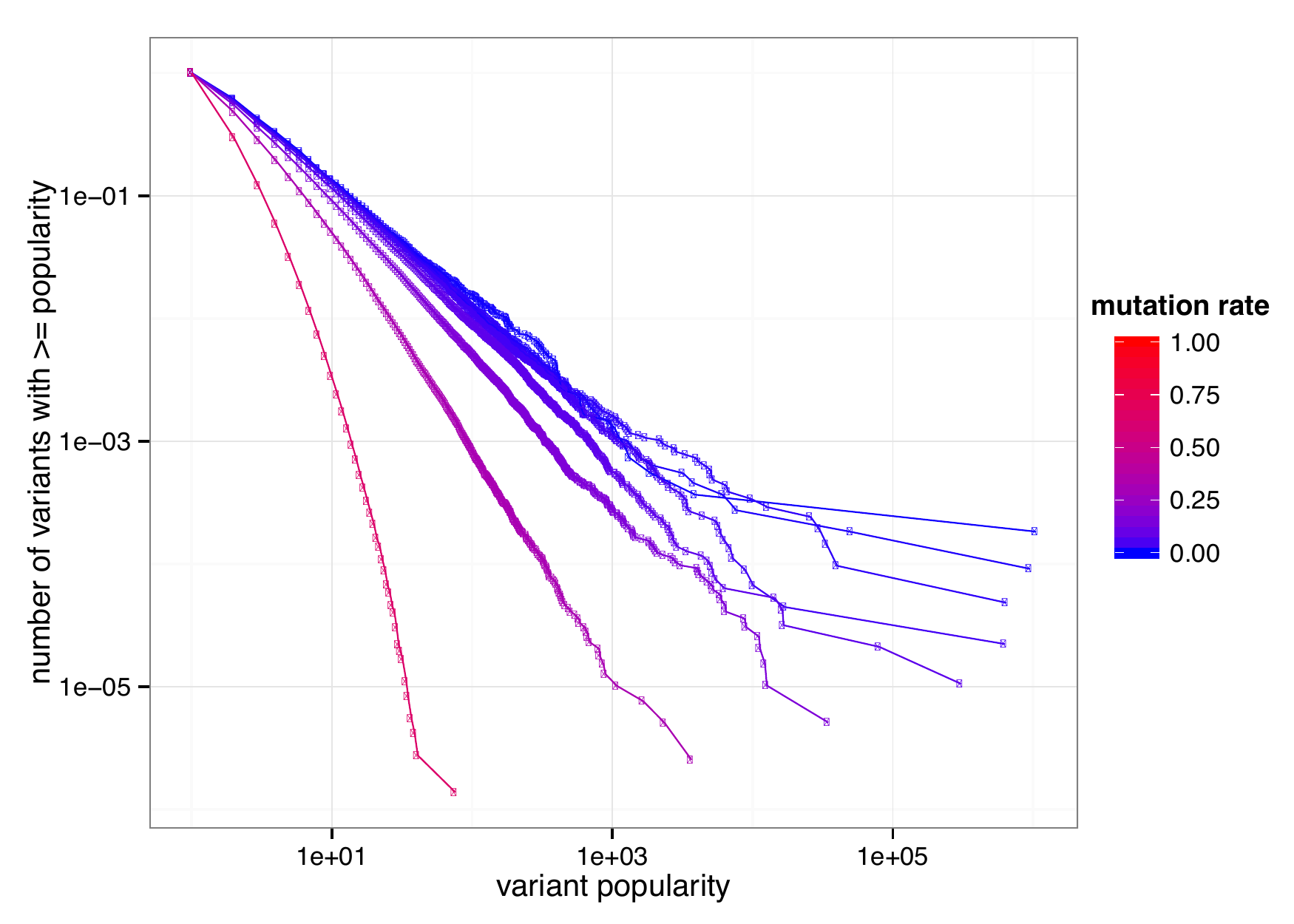} 
\end{tabular}
\caption{{\bf Distributions of variant frequency in (a) a sample of observed memes, (b) a simulated Yule process. }The inset shows the distribution of fitted power-law exponents for all memes with $\geq 10,000$ variants.\label{fig:variantpop}
} 
\end{figure}

The first striking pattern in the data is the uneven popularity of the variants, shown in Figure~\ref{fig:variantpop}. For 121 of the 123 largest memes with over 100,000 distinct variants each, a Kolmogorov-Smirnov test confirmed a power-law fit ($D < 0.05$), with maximum likelihood-fitted exponents of  $2.01\pm0.15$. Expanding to the 435 memes with an excess of 10,000 variants yielded similar exponents: $1.99\pm0.21$.

Although a variety of processes can produce power laws~\cite{newman2005power}, we show that the evolution of memes in an online environment can be modeled by the Yule process~\cite{yule1925mathematical}. This simple process, previously used to model the number of species per genus, bacterial populations~\cite{mandelbrot1974population} and protein domain family sizes~\cite{qian2001protein,koonin2002structure}, contains just two simple components: replication and mutation. The process starts with a single variant of a meme. Each individual instance of the meme has an equal probability of generating a new copy per unit time. The probability of a copy containing a mutation is $\mu$, and we let $r = \frac{\mu}{1-\mu}$ be the ratio between the probability of a mutated and non-mutated copy.

For very large times, the probability that the number of copies of a variant exceeds $y$ is given by~\cite{yule1925mathematical}:
\begin{equation}
Pr(Y \geq y) = r\frac{\Gamma(1+r)\Gamma(y)}{\Gamma(y+1+r)}
\end{equation}
where $\Gamma$ is the gamma function. As the mutation rate $\mu \rightarrow 0$, $r \rightarrow 0$. This simplifies to $Pr(Y = y) = \frac{1}{y (y+1)}$, i.e. a power-law distribution with exponent 2. For higher mutation rates, the tail of the distribution, as $y\to\infty$, is given by
\begin{equation}
Pr(Y \geq y)\sim r \Gamma(1+r) y^{-(1+r)}
\label{eq:approximation}
\end{equation}

Thus the Yule model predicts that memes with a low mutation rate will have variants distributed according to a power-law distribution with an exponent close to 2, which is what we observed with our data  (Figure~\ref{fig:variantpop}(a), inset). The Yule model also predicts that memes with a high mutation rate will deviate from a power law because frequent mutation prevents any single variant from achieving an extremely high number of identical copies. Simulations of the Yule model for a range of mutation rates produce variant popularity distributions, shown in Figure~\ref{fig:variantpop}(b), which bear a close resemblance to the empirically observed distributions in Figure~\ref{fig:variantpop}(a).

\begin{figure}[tbh]
\centering
\includegraphics[width=0.8\columnwidth]{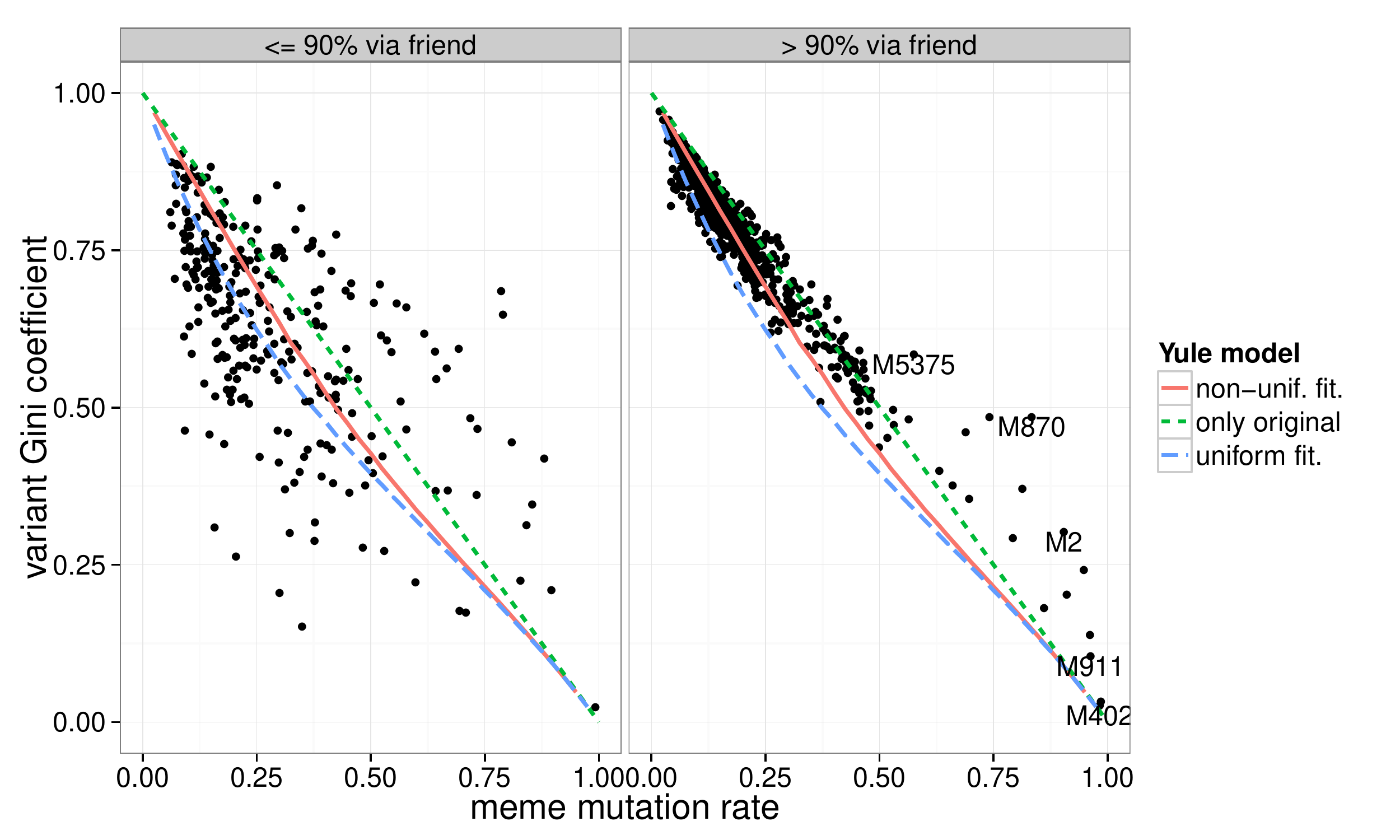}
\caption{{\bf Correspondence between the Gini coefficient $G$ and the mutation rate for memes with over 1000 distinct variants:} (a) 287 memes with $<= 90\%$ of instances being posted following a friend's post, and (b) 876 memes with $> 90\%$ instances posted after a friend. $1-\mu$ is the expected relationship if none of the mutations but the original are viable.  Two simulations, with uniform and variable fitness, are shown for comparison. The identified memes are: M2 (copy sentence from p. 56 of closest book), M402 (child's birthweight in honor of mother's day), M870 (place of birth), M911 (first concert you attended),  M5375 (zodiac sign).\label{fig:ginivsmutation}} 
\end{figure}

The observed power-law distributions for low mutation rates (and their absence for high mutation rates), make the Yule process a plausible mechanism of variant creation and replication. However, many different mechanisms can produce similar distributions, and we want to validate the model predictions further, even for less popular memes where the data is insufficient to confirm a power-law fit. We use the Gini coefficient $G$ to measure the inequality in variant frequency within a meme and to compare it against the $G$ predicted by the Yule model for $\mu$ observed for that meme. $G = 0$ if all variants are equally popular (e.g., if every copy is a mutation), and $G=1$ if only one variant is present (i.e.,  $\mu = 0$). If mutation can occur ($\mu > 0$), but none of the mutated variants are able to produce additional copies, then $G=1-\mu$. For a power-law distribution where  $Pr(Y \geq y) \sim y^{-\alpha}$, $G = \frac{1}{2 \alpha - 1} $\cite{moothathu1985sampling}.  Using Eq.~\ref{eq:approximation}, we can derive the correspondence $G = \frac{1-\mu}{1+\mu}$, valid for $\mu$ close to 0 in the Yule model. This allows us to form a direct prediction from the Yule model of the expected $G$ for a given $\mu$. Since the theoretical prediction holds only for small $\mu$ and uniform fitness, we also simulate the Yule process with both uniform and normally distributed variant fitness, to derive relationships between $G$ and $\mu$ over the full range $0 < \mu < 1$, as shown in Figure~\ref{fig:ginivsmutation}. 

Empirically, we need to be able to determine whether an individual copy is a mutation or not. We do this by connecting meme instances through the friendship graph. If users post a new variant distinct from those previously posted by their friends, the instance is counted as a mutation. If they post an exact copy of one of their friends' previous posts, then it is counted as a copy. Figure~\ref{fig:ginivsmutation} shows a close correspondence ($\rho = -0.97$) between the mutation rate and $G$,  for memes with over 1,000 variants and over 90\%  of instances diffusing through the social network, i.e. having at least 90\% of posts occurring after a friends' post. The curves delineate theoretical and simulated predictions while each point represents an entire meme, with the vast majority of memes falling between these  two curves. This lends additional support that meme propagation follows the Yule model. Note that the correspondence between mutation rate and Gini coefficient is no coincidence and depends crucially on the ability to measure the mutation rate precisely, as demonstrated by the contrast between Figures~\ref{fig:ginivsmutation}(a) and (b). If fewer than 50\% of a meme's posts follow friends' posts of the same meme, it is likely that meme replication is being driven at least partly outside of the social network. This adds uncertainty to the replication path and $\mu$, reducing the correlation between $\mu$ and $G$ significantly ($\rho = -0.60$).

Only a handful of memes mutated at rates above $\mu > 0.5$. Such rates are remarkably high, with more than one out of every two copies resulting in mutation. Indeed, unlike most copy and paste memes which simply instruct the reader to replicate them, memes with high mutation rates additionally specified that they should be mutated by adding specific information: M402 and M4416 (see Table~\ref{tab:memecontent} for meme text) asked for the birthweights of one's children, M6265 for oldest friends, and M431 for billboard charts from one's birthday. We note that it is among this small number of memes mutating at high rates that the Yule model predicts a lower Gini coefficient than is typically observed. A manual examination revealed a consistent explanation: some memes encourage high mutation, but those mutations tend to coincidentally produce the same variant. For example, M870 asks its host to copy it, and add their place of birth. This gives it a very high mutation rate, but also a high $G$, related to the power-law distribution of population center sizes~\cite{gabaix1999zipf,reed2001pareto}, i.e. individuals are likely to independently generate identical strings even while mutating the meme.  Similar coincidences raise $G$ in memes asking the host to copy a sentence from a book (M2), list their zodiac sign (M5375),  and mention the first concert they attended (M911). Therefore, while the Yule model is a good fit for the great majority memes which are ideas evolving ``naturally" under low mutation rates, it is not appropriate for the handful of textual games inducing high mutation rates and coincidental copies. 

Finally, we examine whether mutation rate affects the successful spread of a meme, but find no overall effect of mutation rate on the total number of copies of the meme ($\rho(N_{meme}, \mu_{meme})< 0.1$).  
We further find that meme mutation rates are constant over time ($\rho(t,\mu_{t}) < 0.04$), and that the most popular variants are just as likely to generate mutated copies as less popular ones ($\rho(\text{num mutated copies},\text{popularity}) > .95$), indicating that a meme does not converge to an optimal consensus sequence. 

\subsection*{Mutation characteristics}
While the Yule model explains the overall distributions of different variants, a more detailed look reveals several interesting patterns in the mechanics of how  new variants are created: where edits occur, whether the length matters, and whether an entire meme or just parts of it are inserted into other memes.

\begin{figure}[tbh]
\centering
\includegraphics[width=0.8\columnwidth]{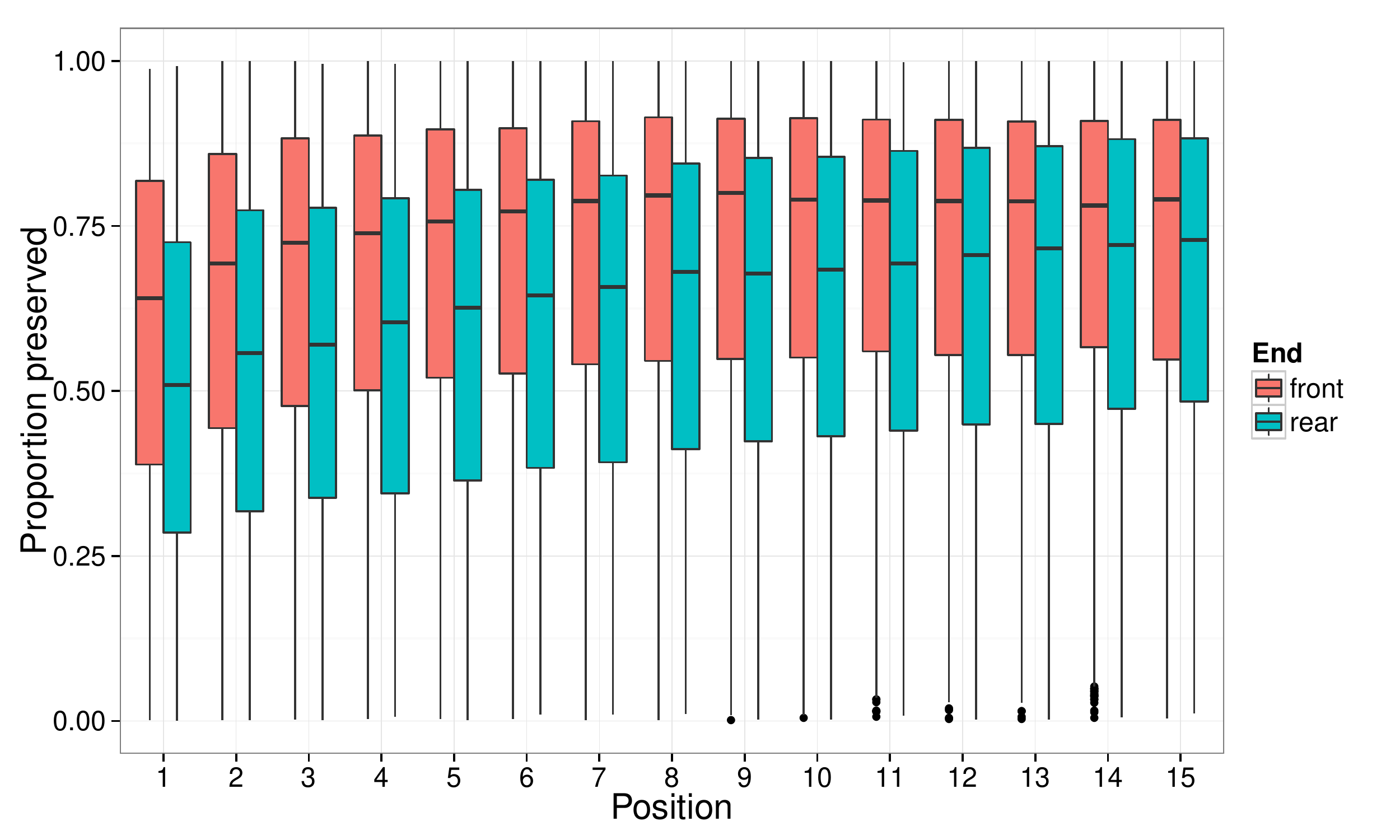} \\
\caption{{\bf Probability that a segment of a meme is preserved as a function of its location within the most popular variant of that meme.} \label{fig:editpreserve}} 
\end{figure}

First, there is a non-uniform probability of mutation along the length of the meme (see Figure~\ref{fig:editpreserve}). This is likely in part due to the copy and paste mechanism for replication, where most users copy and paste the text rather than typing it anew. After pasting, they might prepend, append, or modify the text. To identify the locations where text is most likely to be mutated, we segmented each variant into overlapping sequences of 4 words (4-grams). For each 4-gram within the most popular variant of a meme, we computed the proportion of other variants that contained it. On average, a 4-gram occurring in the middle of the most popular variant is preserved in 70.2\% of other variants. This falls to 59.3\% for the first 4-gram, and 50.3\% for the last.  Edits at the margins are likely to occur because the selection of text to be copied is incomplete, e.g. it misses the start or end. Similarly, pre/post-paste modifications will be at the beginning/end of the meme, unless the user moves the cursor.

\begin{figure}[tbh]
\centering
\includegraphics[width=0.8\columnwidth]{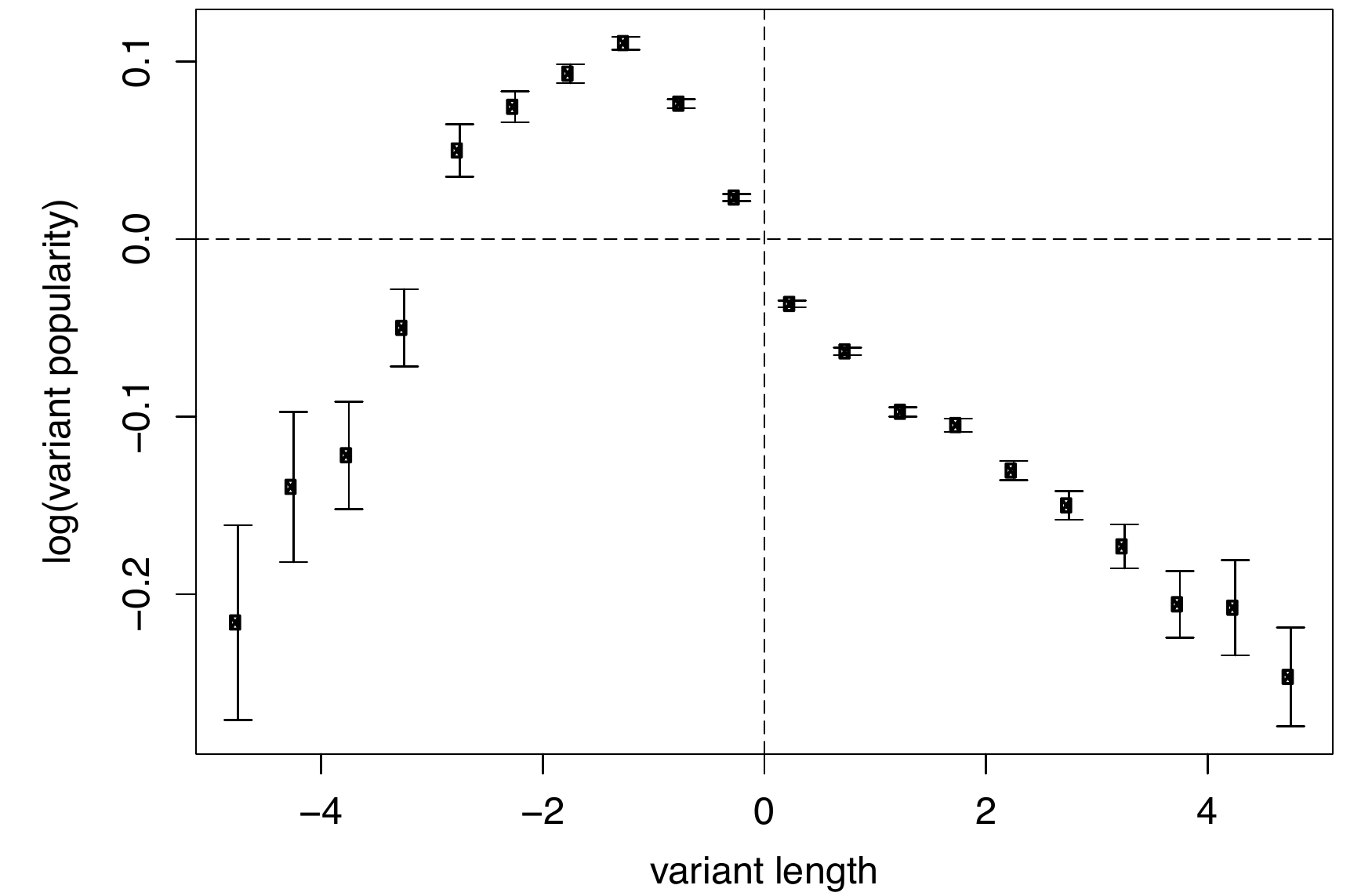}
\caption{{\bf Relative frequency of a variant within a meme as a function of its length relative to other variants.} \label{fig:length}}
\end{figure}

If memes are being replicated through a simple mouse swipe, one would expect a weak selection pressure for shorter variants, reflective primarily of the additional effort of reading and copying a longer meme. Figure~\ref{fig:length} shows just such a weak within-meme selection pressure toward slightly shorter variants. Variants that are too short, and potentially have lost some of the message of the meme, are again not as popular. Consistent with the selection pressure being weak, we observe a high variation both in the average string length between memes (mean=$356.7\pm303.95$), and length of individual variants within a meme (sd=$69.90\pm195.85$).

Many users are exposed to and even participate in replications of several memes. This gives them the opportunity to transfer textual sequences from one meme to another. Just four 4-grams occurred in over 200 memes with a prevalence of between 5 and 95\% of the variants. They contained individual words which were initially used to identify memes in the data, but arranged them in a specific way: ``put this as your'',``copy and paste this'',``this as your status'',``re[-]post if you''.  We computed the ratio $r_g$ of the average frequency of all variants containing a given 4-gram against the average frequency of variants of the same meme that do not have the exact 4-gram. For the four most widespread 4-grams, $r_g > 1.4$, meaning that even though all variants contained replication keywords, these {\it specific} formulations of replication instructions were advantageous.

\begin{table}[h]
\caption{4-grams which conferred replicative advantage on the variants which contained them, within memes where they represented between 5 and 95\% of the population.\label{tab:commonseq}} 
\begin{tabular}{lcc}
4 gram & $r_g$ & \# memes \\ \hline
plain copy and paste & & \\ \hline 
please post this as & 2.59 & 23 \\ 
it as your status & 2.23 & 28 \\
copy and paste and & 2.08 & 20 \\
to copy and paste & 2.04 & 27 \\
this into your status & 2.03 & 29 \\ \hline 
condition/identification/agreement & & \\ \hline
paste if you agree & 2.84 & 25 \\
and paste if you & 2.02 & 35 \\
if you love your & 2.00 & 21 \\
status if you know & 1.98 & 20 \\
your status to show & 1.94 & 28 \\
status to show your & 1.92 & 20 \\
proud to be a & 1.79 & 20 \\
post if you agree & 1.79 & 67 \\
status if you are & 1.79 & 33 \\
if you know someone & 1.78 & 24 \\ \hline
other & & \\\hline
see how many people & 2.85 & 25 \\
status for at least & 1.99 & 29 \\
\end{tabular}
\end{table}

In Table~\ref{tab:commonseq} we list the top 4-grams with the highest $r_g$ and occurring in over 20 memes. Along with copy and paste instructions (``copy and paste'',``as your status'',``into your status''), these include encouragement and allusions to competition (``see how many people''), persistence cues (``status for at least''), or conditions that are easy to match or identify with (``if you love your'', ``if you know someone'', ``paste if you agree'', ``proud to be a''). A specific pattern, ``of you won't'', occurred in  prompts such as  `95\% of you won't copy this, but the 5\% who [have a positive attribute] will'. 144 memes contained at least one variant matching ``won't [\ldots] will''. These variants had significantly higher likelihood of being copied, 10.98 copies on average, relative to an average of 7.05 overall. In 102 of the 144 cases the individual who introduced the phrasing into the meme had either used it previously in another meme, or had a friend who had, suggesting that in a large number of cases the substring could have been transferred from one meme to another. On the other hand, some copy errors, occurring in significant numbers, e.g. ``ago comment like unlike write'', which indicate spurious selection of Facebook boilerplate text below the update, enjoyed only one fourth of the popularity of variants which did not contain them, possibly indicating a notion of {\it grammatical fitness}. This is further supported by a mild negative correlation between the within-meme relative number of English spelling errors $s$ for a variant and its popularity $y$ ($\rho(s,\log(y)) = -0.133, p< 10^{-15}$).

Sometimes, although infrequently, more than a short substring will be shared between separate memes. This occurs when one meme mutates so drastically that it becomes a separate meme, but also when memes recombine at later points, such as when two separate memes are copied and pasted together, thereby merging into a single fused meme. One such example is a variant of an anti-bullying meme (M4038A) with over 300,000 copies, which at one point was pasted together with a meme about a sick child (M4038B). The hybrid, containing most of the text of the two original memes, was exactly copied over 10,000 times.  Similarly, these fused memes contain both messages of the original meme. The combination might not occur for months after the memes are individually introduced. For example, M954, a chain letter meme, was present in fragments since November 2008. The status update character limit imposed at that time did not permit it to be copied entirely, and the fragments posted were likely transferred from copies spreading via email. M26, a similar but distinct chain letter, first appeared in April of 2009. They combined in an awkward copy-and-paste two years later, in October 2011, with over 300 copies made of the combination, including attempts at correcting the copy error in the transition from one meme to the other. 

\subsection*{Adaptation of Memes to Niches}
So far we have presented aggregate characteristics of meme evolution. Examining the evolution of a meme in detail can give insight into how a meme adapts to specific niches within the social network environment. Prior work on cultural transmission has examined how language~\cite{kirby2008cumulative} and music~\cite{maccallum2012evolution} evolve in experimental settings. These lacked the social network structure where different variants are adopted by populations having different preferences. The niche can be time or location-specific; A meme asking mothers to describe their childrens' features, in some variants was prepended by ``In honor of mother's day'' (M4416) and an Amber alert (M246) appeared with the same make of car and license plate but different locations where the child was abducted: Edmonton, KY, Edmonton, Canada, Quebec, ``Washington", and sometimes multiple locations simultaneously.

\begin{figure}[tbh]
\centering
\includegraphics[width=0.8\columnwidth]{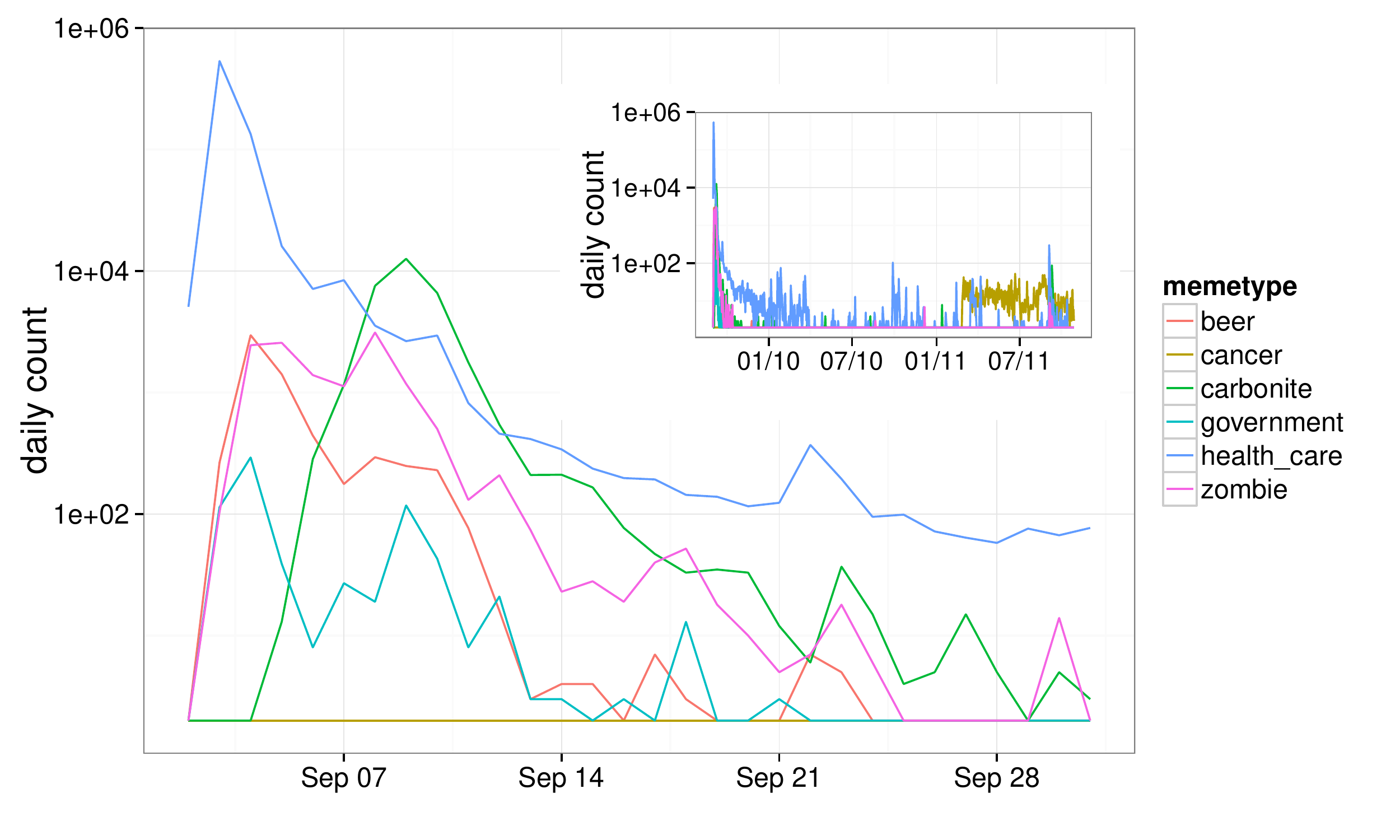}
\caption{{\bf Different variants of the ``no one should\ldots'' meme (see Table~\ref{tab:keywords}) peak in popularity at different times}. The inset shows that the meme persisted in low numbers for two years. \label{fig:nooneshouldtime}}
\end{figure}

\begin{figure}
\centering
\includegraphics[width=0.85\columnwidth]{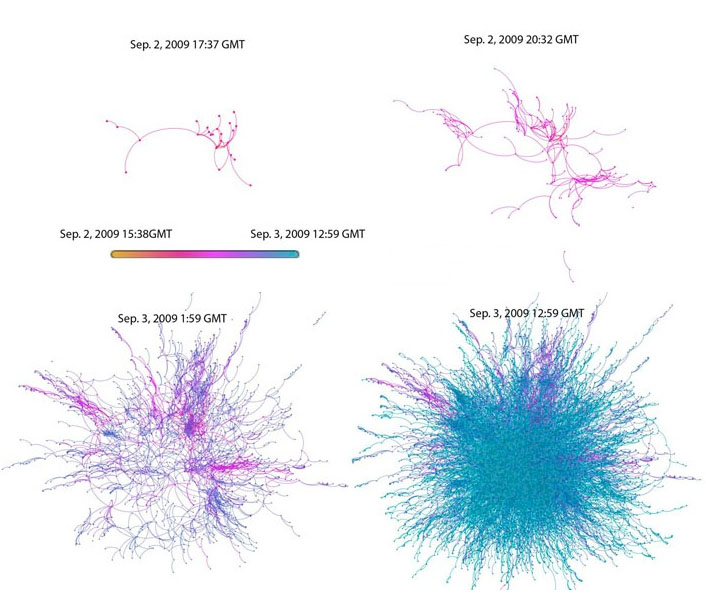}
\caption{{\bf Early diffusion of the ``no one should'' meme.} In a span of two days the meme diffused through many individuals and generated 
diffusion cascades of non-negligible depth.\label{fig:nooneshouldovertime}}
\end{figure}

Here we explore in more detail a meme--``No one should die because they can't afford health insurance\ldots'', whose early diffusion is visualized in Figure~\ref{fig:nooneshouldovertime}. It initially appeared as a political meme but then its variants evolved to appeal to other demographics, including the opposing political party. The original variant can be interpreted as a statement in support of U.S. President Obama's health care reform bill. The meme would appeal primarily to liberals, which we were able to verify by computing the average political leaning for those users who both filled this field out in their profile and posted the meme (see {\it Materials \& Methods} for details).  Although the most prominent variants of the meme propagated primarily from liberal to liberal, some (humorous) variants appealed across the political spectrum ``no one should be without a beer because they cannot afford one,'' while others, criticizing the bill (e.g., ``no one should die because the government is involved with health care\ldots'') appealed primarily to conservatives (see Table~\ref{tab:keywords}). Over time (see Figure~\ref{fig:nooneshouldtime}), the meme had adapted to spread to portions of the network for which the original variant had no appeal.

\begin{table}[h]
\caption{Average political bias of individuals sharing variants of the `no one should' meme (-2: very liberal, +2: very conservative).
\label{tab:keywords}}
\begin{tabular}{l|c|l}
keyword &average bias & example\\ \hline
health care& -0.87& die because they cannot \\
  & &  afford health care\ldots \\ \hline
carbonite  & -0.37 & be frozen in carbonite  because\ldots \\
  & & they couldn't pay Jabba the Hut\\ \hline
zombie  & -0.30 & die  because of zombies if \\
& & they cannot afford a shotgun\ldots \\ \hline
cancer &  -0.02 & have to worry about dying\\
& &  tomorrow but cancer patients do\ldots\\ \hline
wine &  0.15 & go thirsty because they \\
& & cannot afford wine\ldots \\ \hline
beer  & 0.22 & be without a beer because \\
 & & they cannot afford one\ldots \\ \hline
government &  0.88  & die because the government \\
& & is involved with health care\ldots\\ \hline
Obamacare & 0.96 & die because Obamacare \\
& & rations their healthcare\ldots \\ \hline
taxes &  0.97 & go broke because government \\
& & taxes and spends\ldots\\ \hline
\end{tabular}
\end{table}

Once a user posts a meme, they are unlikely to repost it again. That is, exposure to one variant of a meme usually confers immunity to others. However, a small fraction of users post a meme more than once, sometimes years apart, as they identify that the message is still important. In the case of the ``no one should'' meme, 4.36\% of posts were repeat posts by the same individual. Of those posting the meme a second time with a non-healthcare related variant, 40.4\% had posted a healthcare variant previously. Another, more subtle, interaction is that exposure to one variant of a meme can make an individual more susceptible to other variants parodying the original variant. This is evident for the health care meme, where among the 61,979 users posting one of the non-healthcare variants, a full 93.0\% had at least one friend who had previously posted a healthcare variant. This collaborative success of memes has also been observed in a population of image memes~\cite{coscia2013memes}.

\section*{Discussion}
There are convenient analogies between the language of genes and biological evolution that are helpful to us in discussing and analyzing memes. At the most basic level, memes contain two key ingredients that are reminiscent of biological populations: replication and mutation~\cite{elena2003evolution}.  When Richard Dawkins introduced the concept of a meme, he suggested that ``memes should be regarded as living structures, not just metaphorically but technically'' (Richard Dawkins, Selfish Gene p.192.) In this study, we apply this concept more rigorously to Facebook memes and find similarities between genes and memes.  In biology, a gene's genotype  is the DNA sequence, and the gene's phenotype is the resulting function that the gene carries out.  For textual memes, the string is the information that is being passed, or the `genotype.' The `phenotype' is what is expressed by the meme, which can include the meme's message and replication instructions.   

In biology, mutations occur randomly and blindly. Most mutations are neutral, but occasionally, some mutations are deleterious or advantageous~\cite{kimura83neutral}. Neutral mutations do not affect gene function, deleterious mutations are evolutionarily disadvantageous, while advantageous mutations provide a positive benefit to its carriers.  Extending the analogy, neutral edits in memes are minor edits to the string that do not change the meaning of the meme. The Yule process is a neutral model where each meme variant has the same probability of copying itself, and possibly mutating again. Notably, most memes follow the Yule process (Figure~\ref{fig:ginivsmutation}) which suggests that most meme variants are neutral with respect to the parent's meme. We also find some `advantageous' mutations that increase the likelihood of a meme being copied (Table~\ref{tab:commonseq}). There is one critical difference between biology and social media: evolution is a blind process in biology, but in social media, there can be a conscious effort to create mutations that will intentionally spread a meme's presence (i.e. marketing campaigns). We discuss  further similarities, as well as some mechanistic differences, pertaining to the more detailed replication mechanism, below.

The first is the slight preference for shorter variants of the same meme, analogous to the observations of bacteria favoring small genomes for fast replication~\cite{moran2001process}. Both memes and genes are shaped by the same two factors: their length has to be sufficient to encompass information that helps them to replicate, but excessive length slows down replication.

Even more interesting mutation mechanisms also have parallels across both genes and memes. For example, advantageous sequences can occur across multiple memes, likely transferred by a single individual from one meme to another. This process is analagous to lateral gene transfer in bacteria wherein, useful genetic code, e.g. plasmids conferring antibiotic resistance, can be transferred through mechanisms other than replication. Yet another example is that of fusion. In genomes, two single genes are combined in a fusion event to create a single functional gene that retains both functionalities of the two single genes~\cite{enright1999protein}. We observed several fusion events between different memes, some giving rise to popular variants.

Some similarities in the replication mechanism might be spurious. For
example, mutations of both genes ~\cite{chakravarti1999population} and memes occur preferentially
at the boundaries. Many mutations occur at the
beginning and end of a gene (untranslated regions) because most of the
function is in the
middle of the gene and so mutations at the gene boundaries do not
affect gene function~\cite{chakravarti1999population}. Similarly, we observe that meme strings
tend to have text edits at the beginning and end of a meme (Figure~\ref{fig:editpreserve}). The same factor could be at play, e.g. even the most common variants frequently start with a non-essential or generic address: ``everyone", ``attention", and end with words also not essential to the message: ``if you agree", ``takes a second", ``so your fb friends are aware". However, we cannot rule out that the beginnings and ends of memes are simply more vulnerable because of the block copy-and-paste mechanism. 

\textbf{In summary}, we have described how information evolves as it is passed from individual to individual in a social network, sometimes exactly, and sometimes with a modification which produces a new variant. This process is well-described  by the Yule model, with the mutation rate predicting the distribution of popularity among variants. Although many variants appear to emerge from neutral drift, there is evidence of some selection, as successful subsequences were found across memes, and individual meme variants were found to match preferences of individuals transmitting them. A clear limitation of the present study is that the information analyzed was of but one type (copy-and-paste memes) and was observed in a single environment. However, this same environment was the first opportunity to study the evolution of information with precision and on a large scale. We believe our findings are likely to be applicable to a range of environments where information diffuses through multiple steps with a potential for modification, and that the observed evolutionary process carries important implications for the reach and fidelity of information diffusing through social networks.

\section*{Materials and Methods}

Using anonymized data from September 2007 to October 2011, we identified status updates occurring repeatedly across many individuals. Nearly all contained replication instructions such as `copy', `paste', and `repost'. The few exceptions included updates generated automatically by Facebook applications and some ubiquitous memes: jokes and wise sayings. We then collected all status updates containing such English language replication instructions, captured primarily English language variants of memes. Prior to April 2009 textual memes were effectively suppressed , since a 160 character limit left little room for the payload of the meme next to the replication instructions.  \\
\indent Prior to clustering the variants into memes, we removed non-alphanumeric characters and converted the remainder to lowercase. Each distinct variant was shingled  into overlapping 4-word-grams, creating a term frequency vector from the 4-grams. Sorting the meme variants by month, then by frequency, we created a new cluster, i.e. a meme, if the cosine similarity of the 4-gram vector was below 0.2 to all prior clusters. Otherwise, we added the status update to the cluster it matched most closely and adjusted the term-frequency vector of the matching cluster to incorporate the additional variant. We modified the term frequency vector of existing clusters, or created a new cluster, only if the variant frequency exceeded 100 within a month. In a post-processing step we aggregated clusters whose term vectors had  converged to a unigram cosine similarity exceeding 0.4. We then gathered all variants for these 4,087 most significant memes by assigning status updates to them if their cosine similarity exceeded 0.05 using 4-grams and 0.1 using unigrams. The unigram threshold assured that an unrelated status update wasn't erroneously included in a cluster simply for containing a relatively rare substring. \\
\indent 99.49\% of posts identified as memes were made by individual users, allowing us to potentially trace the meme across the friendship graph. The remainder were made primarily by Facebook pages: group entities which enable one-to-many communication. For each meme, we induce the subgraph of directed friendship connections from users who posted the meme to their friends who posted before. We add edges between the {\small $4.89\pm6.30\%$}  posts of the same meme that are from the same user. We sort edges first by Levenshtein edit distance and then by time elapsed, picking the closest match as the parent in the diffusion tree. The process generates a diffusion forest rather than a tree because some origins are unknown, e.g. if a post has been deleted, or if it was propagated through a page rather than an individual's status update.  Picking the textually and temporally closest source gives us a conservative estimate of how much mutation is occurring in a meme.  \\
\indent Our sample includes some memes which never proliferated in large numbers, yet nevertheless produced at least one variant with 100 or more instances. This presented a challenge to estimating the distribution of variant popularity, especially for smaller memes of low mutation rates, where very few distinct variants are generated. We therefore appropriately limited ourselves to memes with a sufficient number of observations to yield accurate statistics. To estimate power-law exponents, which require observations over several orders of magnitude, we included memes with upwards of 100,000 variants. To estimate the required number of variants to generate accurate Gini coefficients, we simulated the Yule process and contrasted the asymptotic Gini coefficient for a meme that had evolved for a long time period, with the $G$ during the early evolution of the meme. We found that the two values matched closely once the meme had grown to over 1,000 variants and so set this as the lower bound for the number of variants for the empirical measurements of $G$. \\
\indent \textbf{Mapping users' political leanings on a scale\label{labelpolitical}}
To understand selection based on political inclinations of individuals in the social network, we used user-provided political affiliations specified in their profiles. Among active users in the United States, approximately 17.6\% of users enter a political affiliation in their profiles. Because of the freeform nature of affiliation entry, there is a long tail of political affiliations, and 1.6 million distinct terms used, from {\it Democratic Party} to  {\it Politics, n: [Poly ``many'' + tics ``blood-sucking parasites'']}  . The top 100 designations account for 86.6\% of all entries.  We limited our analysis to those among the top 100 affiliations that could be mapped on a liberal to conservative scale, -2 being very liberal, 0 being moderate or independent, and 2 being very conservative. We excluded responses such as ``Other'', and ``I don't care'', and ``Libertarian (Party)''. For example, we labeled the Tea Party as very conservative, and the Green Party as very liberal. The full mapping is available  in Table~\ref{tab:leanings}. This included 53 designations, comprising 41.2\% of all users who had entered something into the political affiliation field on their profiles.

\section*{Acknowledgments}
We thank Brian Karrer, Eytan Bakshy, Jon Kleinberg, Jure Leskovec, Alex Dow and Mark Newman for comments. The work was supported in part by NSF IIS-0746646.

\FloatBarrier
\clearpage

\FloatBarrier
\begin{longtable}{p{2cm}|p{12cm}}
\caption{Normalized text of the most common variant of memes mentioned in the article.\label{tab:memecontent}}\\

ID & content \\ \hline
M2 & it s national book week the rules grab the closest book to you go to page 56 copy the 5th sentence as your status don t mention the book post these rules as part of your status \\ \hline
M26 & sorry about this when u r reading this dont stop or something bad will happen my name is summer i am 15 years old i have blonde hair many scars no nose or ears i am dead if u dont copy this just like from the ring copy n post this on 5 more sites or i will appear one dark quiet night when ur not expecting it by your bed with a knife and kill u this is no joke something good will happen to u if you post \\ \hline
M954 & read this isnt fake apparently if u copy and paste this to ten events in the next ten minutes u will have the best day of ur life tomorrow u will either get kissed or asked out if u break this chain u will see a little dead girl in your room tonight in 53 mins someone will say i love you or im sorry or i wanna go out with you \\ \hline
M26 w/ M954  & my name is summer i am 15 years old i have blonde hair and blue eyes i have no nose or ears my body is covered with scars didn t i tell you i m dead my dad killed me with a kitchen knife in the year 2001 if you do not post this on to 10 other pages or groups in the next 15 minutes i will appear tonight by your bed with the kitchen knife that killed me and i will kill you no matter how old you are i will murder you it s up to you if you re post thisor not but this is no lie this is for real now copy this and pasteit onto the walls of 10 other pages or groups your time is runninsee more this isn t fake if you copy and paste this to ten pages in the next ten minutes u will have the best day of ur life tomorrow u will either get kissed or asked out if you break this chain u will see a little dead girl in your room tonight in 53 mins someone will say i love you or im sorry or i wanna go out with you \\ \hline
M402 & to all you mommies out there join the fun and re post this how much did your children weigh at birth \\ \hline
M426 & amber alert edmonton kentucky usa little girl 3 yrs old picked up by man driving grey car license plate quebec 72b 381 canada put this as your status it could save her this kidnapping is recent so do it 3 seconds will not kill you if it were your child what would you want people to do \\ \hline
M431 & keep this going what was the 1 song on the day you were born look it up at joshhosler biz copy and paste with your 1 song \\ \hline
M870 & todays game place of birth everyone please play you will find it interesting to know where your fb friends birth places are copy paste this on your profile then put your place of birth at the end of this sentence \\ \hline
M911 & first concert you attended everyone please play you will find it interesting to know where and what your fb friends first concert was copy paste this on your profile \\ \hline
M1190 & a firefighter is being yelled at for taking too long to get there while trying desperately to save the life of a total stranger holding his bladder because he didn t have time to pee when the alarm sounded starving because he missed one of his three meals tired because the alarm sounded just as he closed his eyes it s now 4 in the morning and missing his family while taking care of yours re post if you are a firefighter love a firefighter or appreciate a firefighter \\ \hline
M4038A & a 15 year old girl holds hands with her 1 year old son people call her a slut no one knows she was raped at 13 people call another guy fat no one knows he has a serious disease causing him to be overweight people call an old man ugly no one knew he had a serious injury to his face while fighting for our country in the war re post this if you are against bullying and stereotyping i bet 95\% of you won t \\ \hline
M4038B & 7yr old with cancer from rosebush mi hi my name is amy bruce i am 7yrs old and i have a large tumor on my brain and severe lung cancer the doctors say i will die soon if this isn t fixed and my family can t pay the bill s the make a wish foundation has agreed to donate 7 for every time this message is sent on for those of you who send this along i thank you so much but for those who don t send it i will pray for you please put this as your status for an hour\\ \hline
M4038 A+B & 15 year old girl holds hands with her 1 year old son people call her a slut no one knows she was raped at age 14 people call another guy fat no one knows he has a serious disease causing him to be overweight people call an old man ugly no one knows he experienced a serious injury to his face while fighting for our country in the war re post this if you are against bullying and stereotyping i bet 88\% of you won t the other 22 aren t heartless and will hi my name is amy bruce i am 7 years old and i have severe lung cancer i also have a large tumor in my brain from repeated beatings doctors say i will die soon if this isn t fixed and my family can t pay the bills the make a wish foundation has agreed to donate 7 cents for every time this message is sent on for those of you who send this along i thank you so much but for those who don t send it what goes around comes around have a heart put this as your status \\ \hline
M4416 & in honor of mothers day post your kids name birthday birth weight birth height and how old they will be this mothers day all proud mommy s re post\\ \hline
M5375 & never give up on 3 people sagittarius aries pisces they are the most adventurous and determined never lose 3 people taurus cancer capricorn they are the most sincere and true lovers never leave 3 people virgo libra scorpio they can keep secrets friendship and they can see your tears the truth never reject 3 people leo gemini aquarius they are true honest friends and always there whats your sign copy and paste in your profile \\ \hline
M5974 & anyone in central texas that has fire fighting experience is strongly encouraged to call 512 978 1187 please copy and paste this and help get the word out \\  \hline
M6265 & aside from blood relatives can you name the person on your friend list who you have known the longest if so copy and paste this as your status along with that person s name \\ \hline

\end{longtable}

\clearpage

\begin{longtable}{l|r}
\caption{Political leanings, ordered by popularity and mapped on a -2 to 2 Liberal to Conservative scale. Prior to December 2010, the responses were limited to a choice of (Very Liberal, Liberal, Moderate, Conservative, Very Conservative, Apathetic, Libertarian and Other). ``Other'' remained a popular choice and accounts for fully 41.4\% of all responses.  ``Liberal'' and ``Democratic Party'' account for 14.8\%, while ``Republican Party'' and ``Conservative'' account for 15.5\%.\label{tab:leanings}} \\
\small
name & valence \\ \hline
Democratic Party & -1 \\
Republican Party & 1 \\
Conservative & 1 \\
Liberal & -1 \\
Moderate & 0 \\
Independent & 0 \\
Very Liberal & -2 \\
Democratic Party & -1 \\
Independence Party of America & 0 \\
Very Conservative & 2 \\
Barack Obama & -1 \\
Conservative Party & 1 \\
Democratic & -1 \\
Democrat & -1 \\
Liberal Democratic Party & -2 \\
Republican & 1 \\
Independent Citizens Movement & 1 \\
Green Party & -2 \\
Republican & 1 \\
Conservatives & 1 \\
Peace and Freedom Party & -2 \\
Tea Party & 2 \\
Constitution Party & 2 \\
OBAMA & -1 \\
Conservative Republicans & 2 \\
Neutral & 0 \\
Socialist Party USA & -2 \\
Independent & 0 \\
Independent Party & 0 \\
Socialist & -2 \\
Middle of the road & 0 \\
Conservative Democrat & -1 \\
Progressive & -1 \\
Moderate Liberal & -1 \\
Moderate Party & 0 \\
Centrist & 0 \\
Conservative Republican & 2 \\
Communism & -2 \\
Rebublican & 1 \\
Conservative Independent & 1 \\
Democratic Party & -1 \\
\hskip 0.2cm of the Virgin Islands &  \\
America First Party & 2 \\
Reform Party of the & 1 \\
\hskip 0.2cm United States of America &  \\
Conservative liberal & -1 \\
Liberal conservative & 1 \\
Fiscal Conservative & 1 \\
Moderate conservative & 1 \\
Liberal Republican & 1 \\
NOBAMA & 1 \\
Communist & -2 \\
GOP & 1 \\
Democrata & -1 \\
OBAMA SUCKS! & 1 \\
\normalsize
\end{longtable}

\end{document}